\documentclass{sf2a-conf2011}
\usepackage{graphicx}
\usepackage{hyperref}
\usepackage[]{natbib}  
\usepackage[cyr]{aeguill}
\usepackage{epstopdf}

\def\BibTeX{{\rm B\kern-.05em{\sc i\kern-.025em b}\kern-.08em
    T\kern-.1667em\lower.7ex\hbox{E}\kern-.125emX}}
\bibpunct{(}{)}{;}{a}{}{,}  

\begin{document}

\TitreGlobal{SF2A 2011}


\title{Testing gravitation in the Solar System with radio science experiments}

\runningtitle{Test of gravity in the Solar Sytem}

\author{A. Hees$^{1,}$}\address{Royal Observatory of Belgium, Avenue Circulaire, 3, 1180 Bruxelles, Belgium}\address{LNE-SYRTE, Observatoire de Paris}

\author{P. Wolf$^2$}
\author{B. Lamine}\address{Laboratoire Kastler Brossel, Paris}
\author{S. Reynaud$^3$}
\author{M.T. Jaekel}\address{Laboratoire de Physique Th\'eorique de l'ENS}
\author{C. Le Poncin-Lafitte$^2$}
\author{V. Lainey}\address{IMCCE, Observatoire de Paris}
\author{V. Dehant$^1$}

\setcounter{page}{237}

\index{Hees, A.}
\index{Wolf, P.}
\index{Lamine, B.}
\index{Reynaud S.}
\index{Jaekel, M.T.}
\index{Le Poncin-Lafitte, C.}
\index{Lainey, V.}
\index{Dehant, V.}


\maketitle


\begin{abstract}
The laws of gravitation have been tested for a long time with steadily improving precision, leading at some moment of time to paradigmatic evolutions. Pursuing this continual effort is of great importance for science. In this communication, we focus on Solar System tests of gravity and more precisely on possible tests that can be performed with radio science observations (Range and Doppler). After briefly reviewing the current tests of gravitation at Solar System scales, we give motivations to continue such experiments. 

In order to obtain signature and estimate the amplitude of anomalous signals that could show up in radio science observables because of modified gravitational laws, we developed a new software that simulates Range/Doppler signals. We present this new tool that simulates radio science observables directly from the space-time metric. We apply this tool to the Cassini mission during its cruise from Jupiter to Saturn and derive constraints on the parameters entering alternative theories of gravity beyond the standard Parametrized Post Newtonian theory.
\end{abstract}

\begin{keywords}
tests of general relativity, radio science
\end{keywords}


\section{Introduction}
Testing General Relativity (GR) is a long standing and worthy effort in the scientific community. From a theoretical point of view, different attempts to quantize gravity or to unify it with other fundamental interactions predict deviations from GR. From an observational point of view, cosmological data can not be explained by the combination of GR and the standard model of particles, requiring the introduction of Dark Matter and Dark Energy. Since these two dark components have not been observed directly, cosmological observations can be a hint that the gravitation theory differs from the Einstein theory of gravity. 

Within the solar system, GR is very well confirmed by different types of experiments: tests of the weak equivalence principle, Post-Newtonian constraints or fifth-force searches. In Sec.~\ref{sec:solarsys}, we briefly recall the solar system constraints on the gravitation theory and we give motivations to go beyond these stringent constraints. In this communication we focus on the possibility to test gravity with radio science measurements (Range and Doppler).   In order to study the impact of alternative theories of gravity on radio science observables, we present a software that simulates Range and Doppler signals directly from the space time metric (and from the initial conditions of the bodies considered). This approach allows one to simulate signals in any metric theory. This software is presented in Sec.~\ref{sec:rangedoppler} as well as some simulations about the Cassini mission.
  
\section{Solar system constraints on gravity} \label{sec:solarsys}    
\subsection{Basis of General Relativity}
General Relativity is built on two main principles. The first one, the Equivalence Principle, gives to gravitation a geometric nature. This principle implies that gravity can be identified to space-time geometry which is described by a metric tensor $g_{\mu\nu}$. Since freely falling test masses follow the geodesics of this metric, their motion is independent of their composition. This universality of free fall can be parametrized by a parameter $\eta$, defined as the relative difference between the accelerations of two test bodies. The universality of free fall has then been tested to an impressive level of $\eta < 10^{-13}$ by Lunar Laser Ranging \citep{williams09} and by torsion pendulum \citep[for a review see][]{adelberger09}. Let us note that even if this principle is one of the best tested in physics some theoretical models coming from unification theories may still produce deviations below the current constraints \citep{damour94}, justifying the necessity to perform more accurate experiments like Microscope \citep{touboul01} or STEP missions \citep{mester01}. 
The Equivalence Principle also defines the behavior of ideal clocks which is independent of their constitution and which measures a geometrical quantity $\tau=\int \sqrt{g_{\mu\nu}dx^\mu/ds \  dx^\nu/ds}\ ds$, a point which is also very well tested.
\\

While the first building block of GR postulates the existence of a metric tensor that determines the trajectory of freely falling bodies (the geodesics) and the behavior of clocks, the second building block concerns the form of the metric tensor. The metric tensor is determined through field equations, which in the case of GR, are the Einstein equations 
\begin{equation}
	G_{\mu\nu}=R_{\mu\nu}-\frac{1}{2}g_{\mu\nu}R=\frac{8\pi G}{c^4}T_{\mu\nu},
\end{equation}
$G_{\mu\nu}$ is Einstein curvature tensor, $R_{\mu\nu}$ and $R$ are Ricci and  scalar curvatures (these are derived from the metric), $G$ is Newton constant, $c$ the speed of light and $T_{\mu\nu}$ the stress-energy tensor. These equations allow one to determine the metric tensor from the energy and matter contents of space-time.
\\

If one considers the solar system and only the Sun's gravitational contribution, the metric tensor can be written as an expansion of the Newton potential $\phi_N=-\frac{GM}{rc^2}$ (where $M$ is the mass of the central body and $r$ a radial coordinate)
\begin{equation}      \label{metricSun}
	ds^2=(1+2\phi_N+2\phi_N^2+\dots)c^2dt^2-(1-2\phi_N+\dots)d\vec x^2.
\end{equation}
This expansion is justified since the value of the Newton potential in the solar system is always smaller than $\phi_N<10^{-6}$. The metric (\ref{metricSun}) is written in isotropic coordinates. 
\\

The previous two principles produce important effects that need to be taken into account for high precision astrometry or for space missions: first, the dynamic is different from Newton gravity (one major consequence is the advance of planet perihelia) ; then, the propagation of light is affected ; finally clocks measure a proper time which is different from coordinate time $t$ appearing in the metric. All these effects need to be modeled for space missions in addition to other classical effects. As a consequence, space missions constitute an excellent laboratory to test gravity.

\subsection{Solar system tests of General Relativity}
Gravity tests allow one to compare measurements with the predictions made by some theoretical framework. Within the solar system, two important frameworks have been developed: the Parametrized Post Newtonian formalism fully described in \cite{will93} and the fifth force search fully described in \cite{talmadge88,adelberger09}. Both formalisms embed GR metric within a wider class of specifically parametrized metric. 
\\

The PPN formalism extends the metric (\ref{metricSun}) by introducing Post Newtonian Parameters in expansion (\ref{metricSun}). In the simplest case, two parameters are introduced and the metric describing a spherical and static sun becomes
\begin{equation}
		ds^2=(1+2\phi_N+2\beta\phi_N^2+\dots)c^2dt^2-(1-2\gamma\phi_N+\dots)d\vec x^2.    
\end{equation}       
Testing GR with PN formalism consists in measuring the PPN parameters $\gamma$ and $\beta$ and comparing the experimental results with their GR value ($\gamma=\beta=1$). Thirty years of precise experiments have constrained PPN parameters very closely around GR \citep[for a review, see][]{will06}. In particular, the observation of the Shapiro delay of the Cassini probe during a solar conjunction in 2002 gives the best constraint on the $\gamma$ PPN parameter known to date \citep{bertotti03}
\begin{equation}      \label{gamma}
	\gamma-1=(2.1\pm2.3)\times10^{-5} .
\end{equation}
This constraint is confirmed by deflection measurement with VLBI \citep{lambert09} or tracking Mars orbiter \citep{konopliv11}. The present best constraint on the $\beta$ parameter comes from the solar system ephemerides (assuming the $\gamma$ parameter to be given by (\ref{gamma}))  \citep{fienga11}
\begin{equation}
	\beta-1=(-4.1\pm7.8)\times10^{-5}.
\end{equation} 
Other type of experiments also confirmed this constraint like Lunar Laser Ranging \citep{williams09} or Mars orbiter tracking \citep{konopliv11}. 
\\

The other type of formalism often used in solar system tests of gravity, the fifth-force search, consists in a search of the dependence of the Newton potential $\phi_N$ with the radial coordinate. This formalism parametrizes deviations from the Newtonian potential with a Yukawa potential, justified by unification models. More precisely the scale dependance of the Newton potential modifies the temporal part of the metric
\begin{equation}
	g_{00}=1+2\phi_N\left(1+\alpha e^{-r/\lambda}\right)+2\phi_N^2=[g_{00}]_{GR}+2\phi_N \alpha e^{-r/\lambda}.
\end{equation}                                                                                                
The $\alpha$ parameter characterizes the amplitude of the deviation with respect to the Newtonian potential and the $\lambda$ parameter is a range related to the mass of the new particle that would mediate this fifth interaction. These parameters have been tested in a very wide range \citep[see Fig. 31 of][]{konopliv11}. In particular, the $\alpha$ parameter is constrained to a very high level of accuracy ($\alpha<10^{-10}$) at  Earth-Moon and Sun-Mars distances. From this picture, we can also see that windows remain open at very short distances and at very large distances.

\subsection{Is it necessary to go beyond these tests ?}\label{sec:beyond}
From the arguments presented in last section, one may wonder if it is necessary to continue to go beyond the present constraints on the gravitation theory. The answer is positive for several reasons. Firstly, there exist theoretical models predicting deviations smaller than the current constraints. For example, a tensor-scalar theory of gravity can naturally be attracted towards GR by a cosmological mechanism~\citep{damour93} and lead to a deviation of the $\gamma$ parameter smaller than the Cassini constraint. Another example is given by the chameleon field \citep{khoury04} where deviations of GR are hidden in region of high density (in the solar system) and remain smaller than current constraints.  

Then, it appears interesting to search for deviations in regions where no test has been performed so far. For example, a fifth force may be searched at very small range or at very large range where stringent constraints are missing. Let us mention that not all alternative theories of gravity are entering the PPN or fifth-force framework. It would be instructive to extend these frameworks to include new types of deviations. To illustrate this, we consider two alternative theories of gravity that do not enter the PPN or fifth-force formalism.

Such a first alternative theory of gravity is provided by \textit{Post-Einsteinian Gravity (PEG)} \citep{reynaud05,jaekel05,jaekel06,jaekel06b}. This theory is based on a non local extension of Einstein field equations, as suggested by radiative corrections. Phenomenologically in the solar system, the space time metric can be parametrized by two radial dependent potentials $\delta\Phi_N(r)$ and $\delta\Phi_P(r)$
\begin{eqnarray}
	g_{00} &=  & 1+2\phi_N+2\phi_N^2+2\delta\Phi_N  \\
	g_{ij} &=& \delta_{ij}\left(-1+2\phi_N+2\delta\Phi_N-2\delta\Phi_P\right).
\end{eqnarray}
This parametrization extends the PPN formalism by replacing parameters by functions.

The second example considered in this communication concerns a particular effect due to MOND theory. In \cite{blanchet11}, it is shown that MOND induces within the solar system an \textit{External Field Effect}. This effect is modeled by a quadrupolar contribution to the Newtonian potential
\begin{equation}
	\phi_N=-\frac{GM}{rc^2}-\frac{Q_2}{2c^2}x^ix^j\left(e_1e_j-\frac{1}{3}\delta_{ij}\right)
\end{equation}
$e_i$ is a unitary vector pointing towards the galactic center and $2.1 \ 10^{-27} \ s^{-2}\leq Q_2\leq 4.1\ 10^{-26}\ s^{-2}$ is the value of the quadrupole moment whose value depends on the MOND function.

These examples show that there are motivations to improve current tests of gravity but also to look at situations previously untested. This can be done by considering theories not entering the traditional frameworks or looking in  regions of parameters where strong constraints are missing.

\section{Range and Doppler simulations}\label{sec:rangedoppler}
In this work we focus on the possibility to perform gravity tests with radio science measurements. This kind of test has already been very successful when deriving the PPN $\gamma$ constraint with Cassini (see previous section). In order to provide estimation of the order of magnitude and of the signature that an alternative theory of gravity produces on radio science signals (Range and Doppler), we have developed a software that simulates these signals. In order to allow for a wide class of alternative theories of gravity, Range and Doppler are simulated directly from the space-time metric $g_{\mu\nu}$. This means that it is very easy to change the gravitation theory by changing the metric.
\\

Since GR and the alternative theories of gravity we consider are covariant, we are free to choose the coordinates to work with (gauge freedom). On the other hand, observations are covariant quantities (or gauge independent). Therefore, it is very important to produce simulations that are also covariant. In radio science, covariant quantities are based on proper time (time given by ideal clocks). The Range is defined as the difference between the reception proper time and the emission proper time and the Doppler is defined as the ratio of the received proper frequency over the emitted proper frequency. Three different steps are needed to simulate these quantities: the derivation and integration of the equations of motion, the derivation and integration of the equations of proper time and the computation of time transfer in curved space-time. These three steps are computed from the space-time metric by methods described in \cite{hees11}. Finally in order to investigate the observables of an alternative theory of gravity in the Range and Doppler data we perform a least-squares fit in GR on the different parameters (initial conditions and masses of the bodies) and search for identifiable signatures in the residuals. 
\\

In the following, as an example, we present simulations of a two-way Range and Doppler signals for Cassini spacecraft from June 2002 during 3 years (when the probe was between Jupiter and Saturn). A simplified model is built with the Sun, the Earth and Cassini spacecraft. We successively consider Post-Einsteinian Gravity and MOND External Field Effect.

\subsection{Post-Einsteinian Gravity (PEG)}
As a simplified preliminary study, we focus on the effects of the potential $\delta\Phi_P(r)$ on the radio science signals. Indeed, the potential $\delta\phi_N$ is already very tightly constrained (from fifth force measurement). We consider a series expansion of this potential, that is to say we suppose the spatial part of the metric to be modified as
\begin{equation}
	g_{ij}=\left[g_{ij}\right]_{GR}-2\delta_{ij}\left(\chi_1 r+\chi_2r^2+\delta\gamma\frac{GM}{c^2r}\right)
\end{equation}  
$M$ is the Sun mass, $\delta \gamma=\gamma-1$ is related to the PPN parameter and $\chi_1$ and $\chi_2$ are parameters characterizing deviations due to linear and quadratic terms in the metric.

\begin{figure}[ht!]
 \centering
 \includegraphics[width=0.45\textwidth,clip]{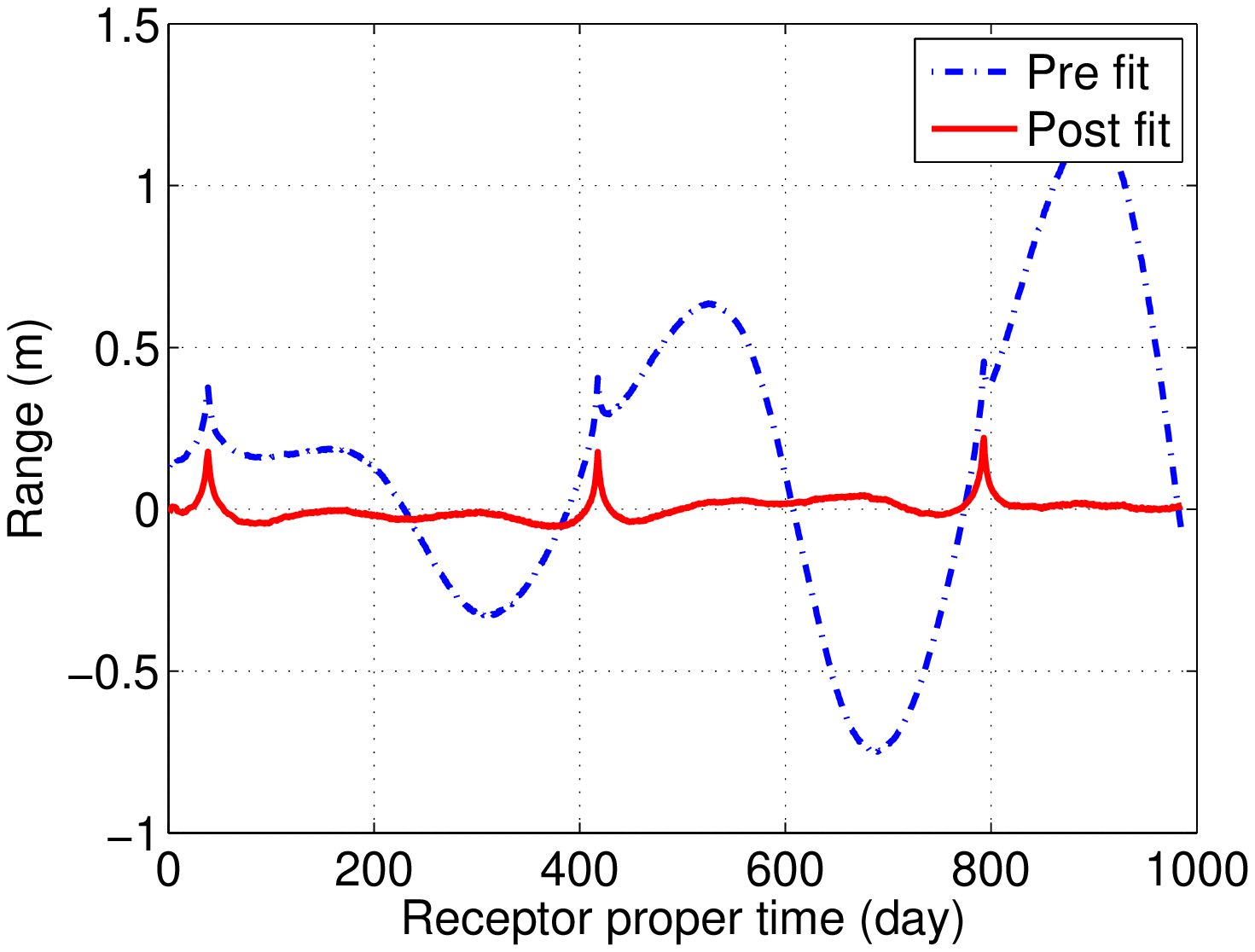}      
 \includegraphics[width=0.45\textwidth,clip]{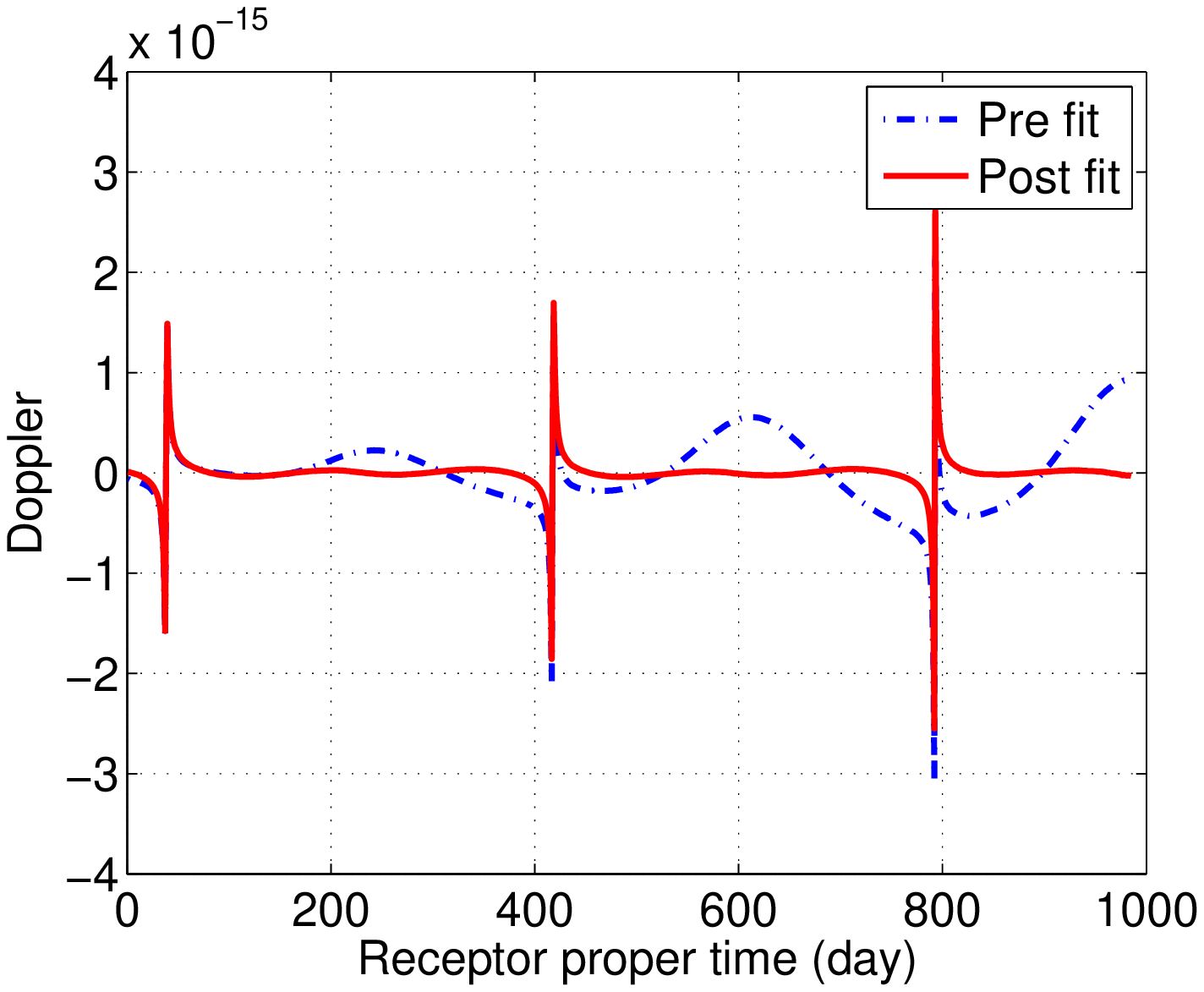}      
  \caption{Representation of the Range (left) and Doppler (right) signals due to a PEG theory of gravity with $\delta\gamma=\gamma-1=10^{-5}$. The blue (dashed) lines represent the direct difference between simulations in GR and simulations in the alternative theory of gravity (with the same parameters). The red (continuous) lines represent the residuals after the least-square analysis (the observable signals).}
  \label{fig1}
\end{figure}

We performed different simulations with different values for the three PEG parameters. For example, Fig.~\ref{fig1} represents the Range and Doppler residuals due to the presence of a PPN $\gamma$ deviation of $\delta\gamma=\gamma-1=10^{-5}$. The three peaks occur at solar conjunctions. The signal due to the conjunctions is not absorbed at all by the fit of the initial conditions. The simulated data are comparable with real data obtained in \cite{bertotti03}. 
                                                                                                                         
\begin{figure}[ht!]
 \centering
 \includegraphics[width=0.325\textwidth,clip,trim=5 0 20 15]{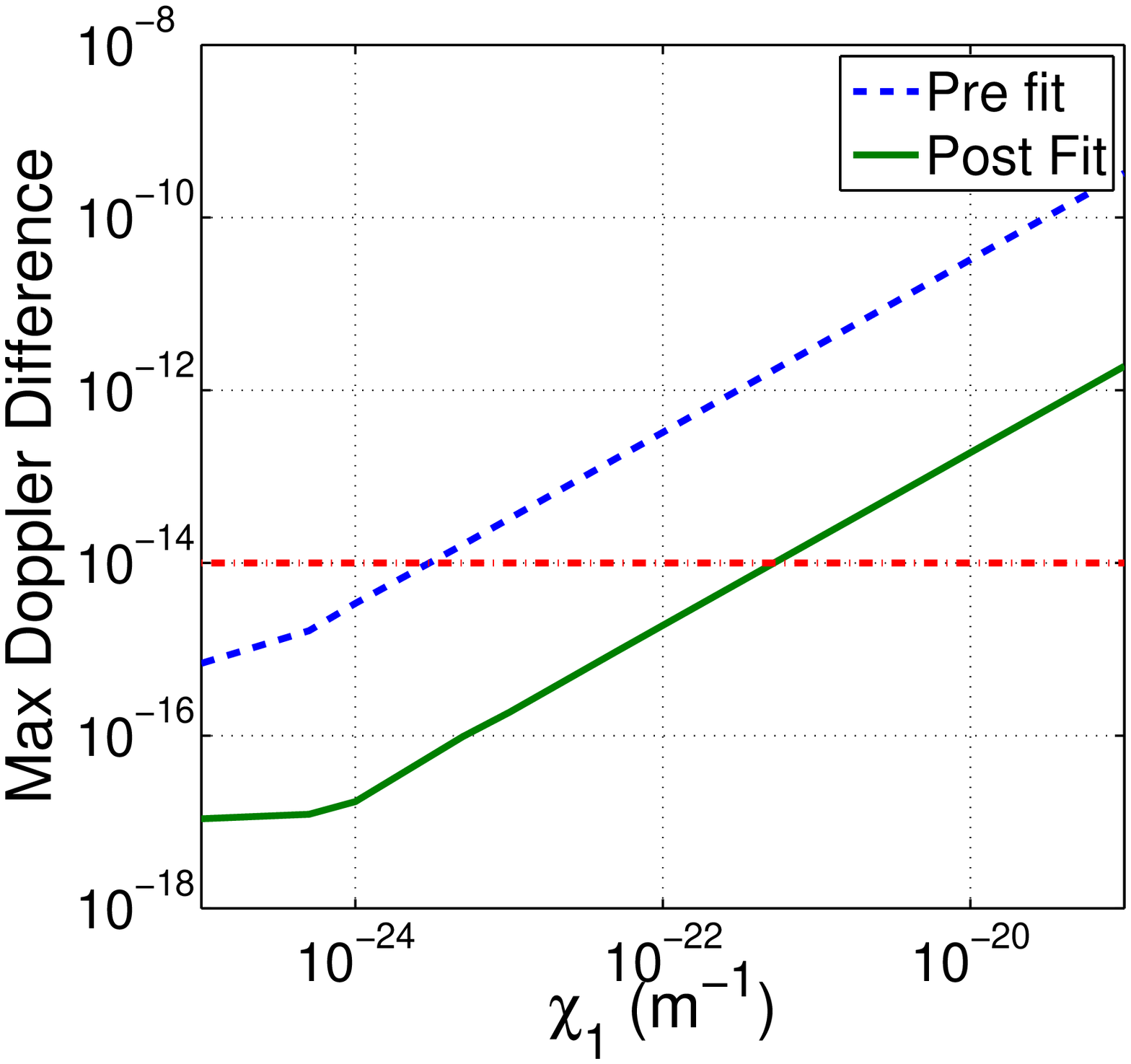}
 \includegraphics[width=0.325\textwidth,clip,trim=5 0 20 15]{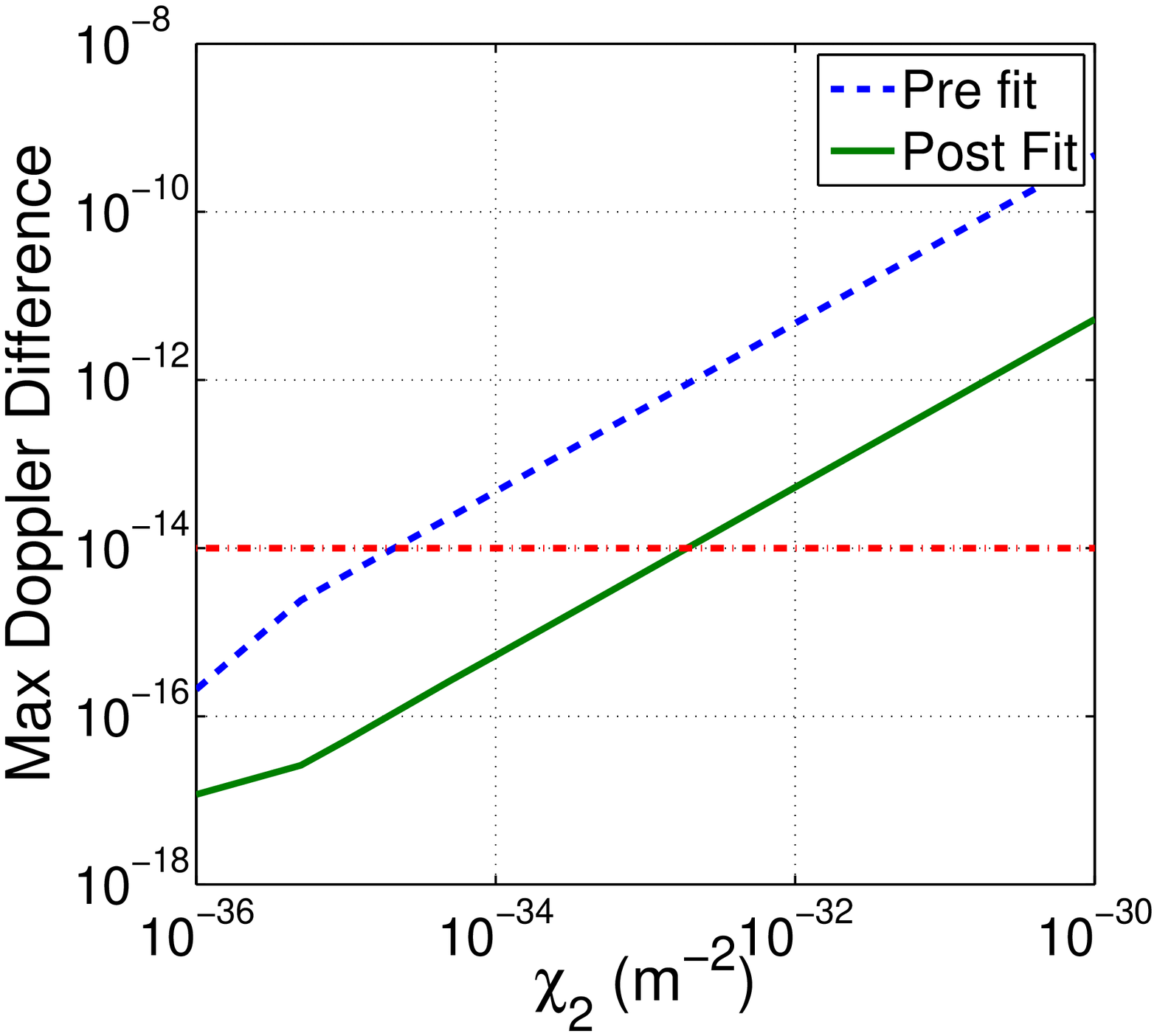}      
 \includegraphics[width=0.325\textwidth,clip,trim=5 0 20 15]{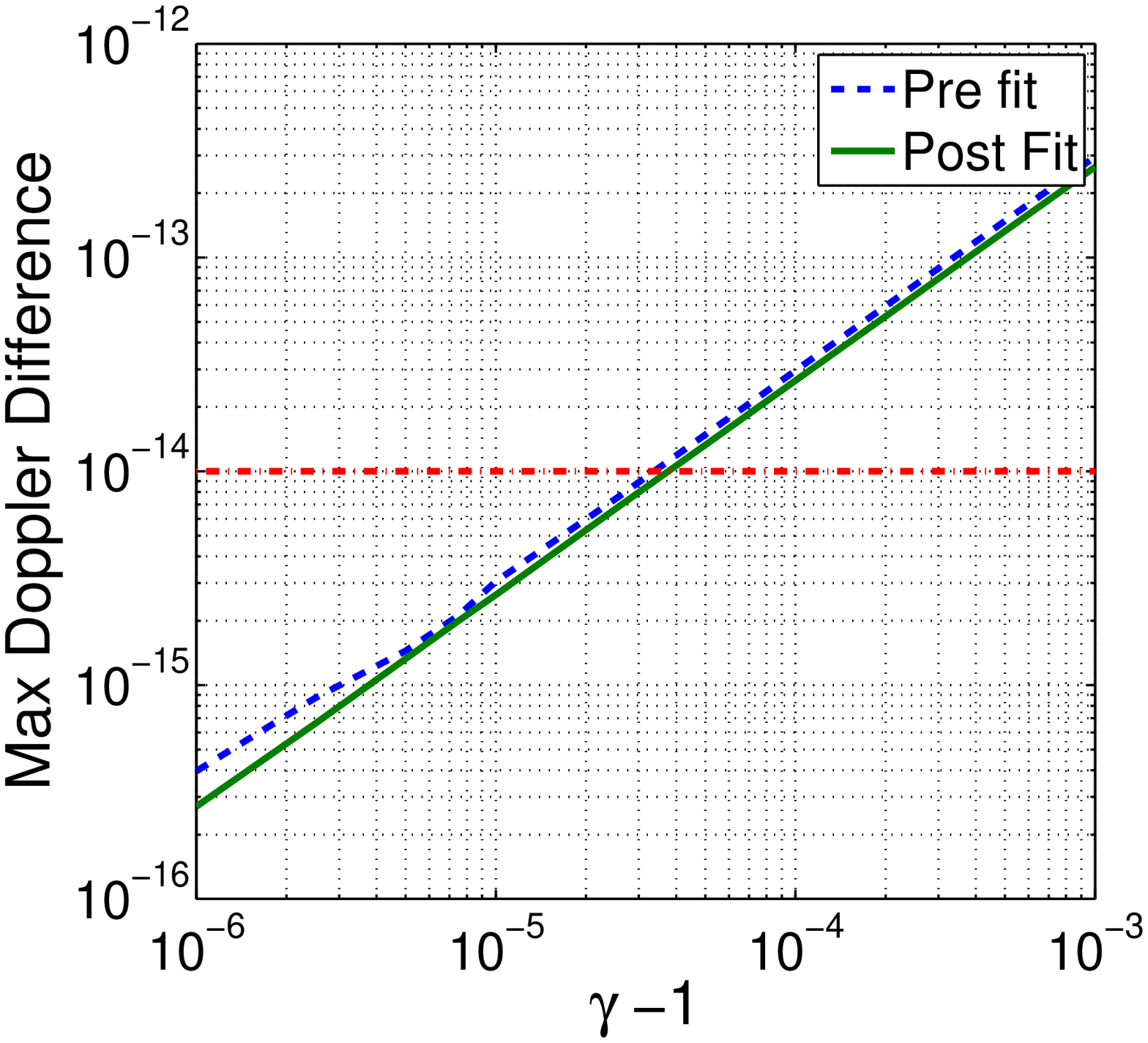}      
  \caption{Representation of the maximal Doppler signal due to PEG theory (parametrized by the three parameters $\chi_1$, $\chi_2$, $\delta\gamma$) for the Cassini mission. The blue (dashed) lines represent the direct difference between the Doppler in alternative theories and GR (with the same parameters) while the green (continuous) lines represent the maximal residuals that can be observed after the least-square analysis. The red lines represent the Doppler Cassini accuracy.}
  \label{fig2}
\end{figure}

To summarize the simulations performed, Fig.~\ref{fig2} represents the maximal Doppler residuals due to PEG theories of gravity (parametrized by their values of $\chi_1$, $\chi_2$ and $\delta\gamma=\gamma-1$). By requesting the maximal residuals to be smaller than the Cassini precision on Doppler signal ($10^{-14}$), we can estimate an upper limit for the PEG parameters: $\chi_1<10^{-23}\ m^{-1}$, $\chi_2<2\ 10^{-33}\ m^{-2}$ and $\delta\gamma<3\ 10^{-5}$ (which is very similar to the real estimation (\ref{gamma})).

\subsection{MOND External Field Effect}  
\begin{figure}[ht!]
 \centering
 \includegraphics[width=0.45\textwidth,clip]{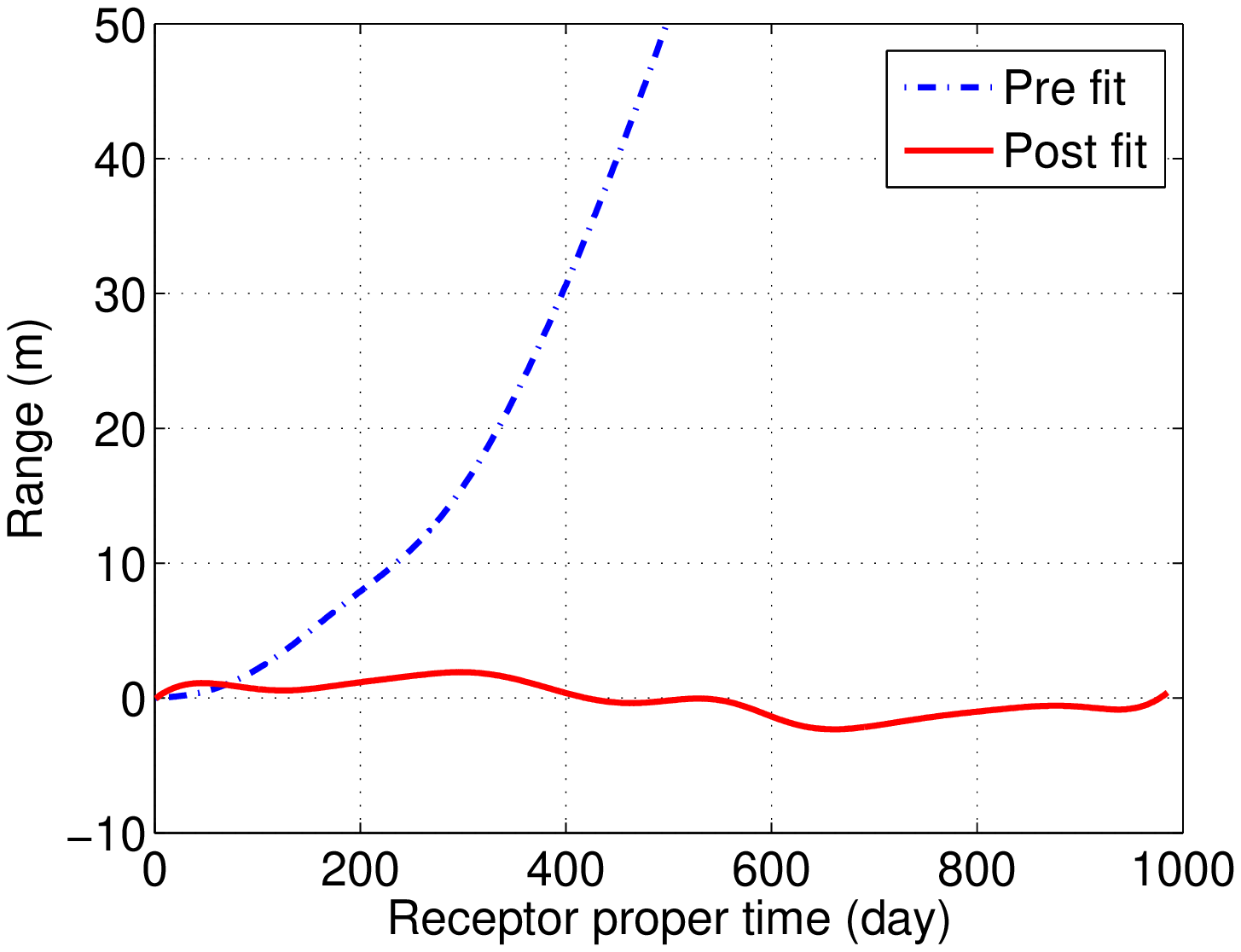}      
 \includegraphics[width=0.45\textwidth,clip]{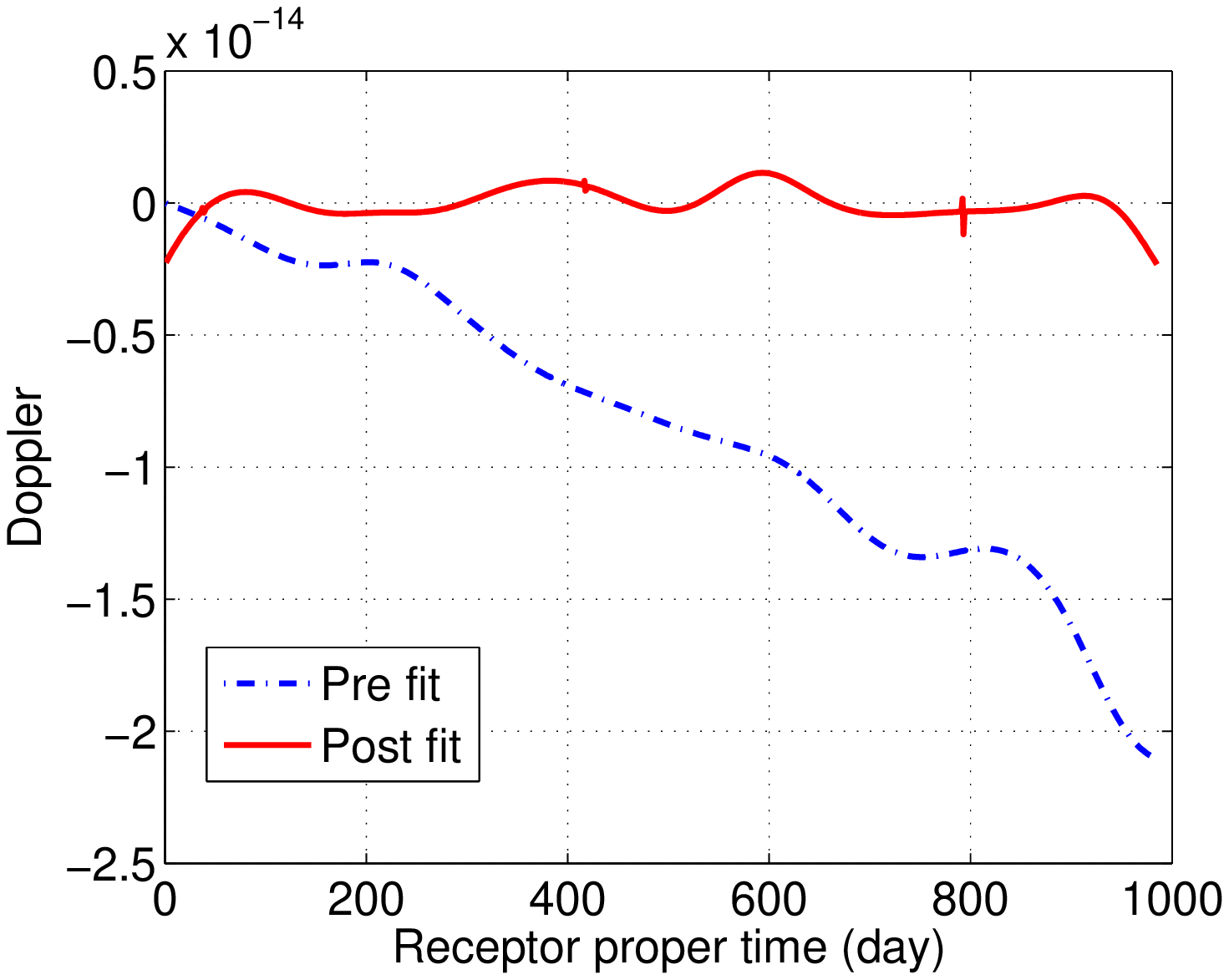}      
  \caption{Representation of the Range (left) and Doppler (right) signals due to the MOND External Field Effect with $Q_2=4.1 \ 10^{-26} \ s^{-2} $. The blue (dashed) lines represent the direct difference between simulations in GR and simulations in the alternative theory of gravity (with the same parameters). The red (continuous) lines represent the residuals after the least-square analysis (the observable signals).}
  \label{fig3}
\end{figure}

In the framework of MOND theory, \cite{blanchet11} have shown that there exists an effect in the Solar System, the External Field Effect (EFE), which is modeled by a quadrupole contribution (see Sec.~\ref{sec:beyond}). Results of our simulations for this effect are given in Fig.~\ref{fig3} where Range and Doppler residuals due to the EFE with $Q_2=4.1 \ 10^{-26}\ s^{-2}$ (the maximal value allowed by the theory) are represented. The residuals are below the Cassini accuracy ($10^{-14}$ in Doppler). Therefore, the Cassini arc considered here appears to be insufficient to provide a satisfactory test of MOND EFE. Note that this may be improved by considering other arcs or a dedicated space mission.

\section{Conclusions}
Starting from the basis of GR, we have presented current gravity tests performed in the Solar System. We have given motivations to increase the accuracy of the current constraints and to look at situations previously untested. In this context, we have presented a new software that performs radio science simulations from the space-time metric. With this tool, we are able to simulate any space mission in any alternative metric theory of gravity and to give the signature of an hypothetical alternative theory of gravity on the Range and/or on the Doppler. As examples, we have presented simulations of the Cassini mission in PEG theory of gravity and with the MOND External Field Effect. For the PEG theory, constraints have been derived for the parameters entering the expression of the spatial part of the metric. For the MOND EFE, we have shown that the predicted effect is too small to be detected with the considered arc of the Cassini mission.

In the future, further simulations can be done for other theories and other (future and past) space missions to answer the following question: can a particular theory of gravity be observed with Range/Doppler measurements of a specific mission ?

\begin{acknowledgements}
A. Hees is research fellow from FRS-FNRS (Belgian Fund for Scientific Research) and he tanks FRS-FNRS for financial support.
\end{acknowledgements}

\bibliographystyle{aa}  
\bibliography{hees} 

\end{document}